# DIAL: Distributed Interactive Analysis of Large Datasets

D. L. Adams

*Brookhaven National Laboratory, Upton NY 11973, USA*

DIAL will enable users to analyze very large, event-based datasets using an application that is natural to the data format. Both the dataset and the processing may be distributed over a farm, a site (collection of farms) or a grid (collection of sites). Here we describe the goals of the project, the current design and implementation, and plans for future development. DIAL is being developed within PPDG to understand the requirements that interactive analysis places on the grid and within ATLAS to enable distributed interactive analysis of event data.

## 1. INTRODUCTION

Modern HEP (high energy physics) experiments collect, generate and reconstruct tremendous amounts of data. LHC [1] experiments are expected to produce 1-10 pB each year for tens of years. In principle, the data for each experiment is available to all scientific members of the collaboration which builds and operates the detector. In practice, it is a challenge to provide data access to the more than 2000 globally-distributed physicists that collaborate on each of the largest experiments.

Few, if any, scientists or institutions can afford to store or process all the data for one of these experiments. Instead, the storage and processing are distributed throughout the collaboration. The initial processing, called production, is done in a well-regulated manner and the results are shared by all. Final physics measurements and direct searches for new physics are done by individuals or small groups of physicists in a more chaotic process known as analysis.

A typical analysis uses a small fraction of the production results as input and them performs selections and other data manipulations to arrive at a conclusion about the existence or properties of a physical process. Much of the analysis is done interactively, i.e. the scientist submits a request to process a dataset, receives a response in seconds or minutes, and then submits another request. The size of the input dataset is often limited by the requirement of interactive response.

For large datasets, the response time can be improved by distributing the processing. This helps in two ways. First, it enables multiple processors to work on the problem at the same time. Second, because the data itself is distributed, there is the possibility to send the process to the data rather than moving the data. Thus, distributed processing allows more physicist queries over larger datasets. This decreases the likelihood that features of the data will be overlooked and enhances the discovery potential of an experiment.

### 1.1. Grid computing

The emerging computing grid [2] infrastructure promises to standardize the worldwide distribution of both data and processing. This standardization will make it easier for diverse users to contribute to and make use of a common pool of storage and processing resources. Experiments adopting the grid model can consequently expect more resources to be available to their collaborations. They can also expect a fairer allocation of the use of these resources. Both the experiment as a whole and the individual contributors will have a voice in this allocation. It will also be easier to share resources with other HEP experiments and with non-HEP activities.

Although most of the work so far done on computing grids assumes a batch-oriented mode of operation, the extension to interactive use is very attractive. An interactive user typically makes a request and then pauses to study this result before generating the next request. This leads to low average resource usage but high peak demand while the user waits for a response. The large resource pool available on a grid is a natural match. The compute cycles between requests can be consumed by other interactive or batch users.

## 2. GOALS

DIAL (Distributed Interactive Analysis of Large datasets) [3] is a project to investigate HEP distributed interactive analysis. It has three primary goals:
1. Demonstrate the feasibility of distributed analysis of large datasets
2. Set corresponding requirements for grid components and services
3. Provide the ATLAS [4] experiment with a useful distributed analysis environment

The first goal can be restated as evaluating how large a dataset can be analyzed interactively. This includes limits imposed both by the available resources and by the scalability of the distributed computing model.

The second reveals our expectation that the data and processing will be distributed over a computing grid and that scaling to the largest datasets will take advantage of this. This work is done in as part of the interactive analysis group (CS-11) of the PPDG (Particle Physics Data Grid) collaboration [5].

The third goal is intended to keep us in the real world. ATLAS is one of the large LHC experiments and if the model we develop is not applicable there, then it is likely not of much value. It would be useful to add another experiment to demonstrate that the model is generic (see the following).

The first and last goals imply that we will deliver a concrete environment for distributed analysis; not just a proposal or design. This environment will be generic and it will, to the greatest extent possible, be layered on top of existing or expected components and services.

By generic, we mean that the bulk of what we develop or identify can be used by different experiments with minimal constraints imposed on the format in which the data is stored. We will not develop the front-end analysis environment but will provide a system that is easily integrated into and accessed from existing and future environments. Ideally, the model will not be restricted to HEP.

DIAL is identifying the components and services required for distributed processing, especially those relevant to the grid. It makes use of existing components where they exist and expresses requirements where they do not. In the latter case we will look to other projects to deliver the required functionality although we may deliver simple prototypes to cover the period before these components appear.

## 3. EVENT PARALLELISM

One important assumption is made about the data. It is assumed to be organized into events that are processed independently. When a dataset is processed, each event in the dataset is processed in the same way and the results generated by the processing of each event are concatenated to produce the overall result. The value of the latter does not depend significantly on the ordering of the processing.

This structure of the data allows for event parallelism where different events are processed in different jobs which may be run on different compute nodes, different types of nodes, and even at different sites.

### 3.1. Outside HEP

Realms outside of HEP may not use the word event but often have an equivalent record which forms an analogous processing unit. Even within HEP, one may wish to choose a different processing unit such as tracks, jets or calibration or alignment measurements. Our primary interest is HEP event data but the DIAL model applies equally well to these other realms with the word event interpreted accordingly.

This generalization is important to keep in mind especially in our grid discussion because we would like to identify components and services with a wide range of applicability.

## 4. DESIGN

DIAL uses event parallelism to distribute the processing of a dataset over multiple jobs. A user working inside an interactive analysis environment gives DIAL a request to process a dataset in a given response time. DIAL then creates one or more processing jobs, concatenates their results and returns the overall result to the user. This structure is illustrated in figure 1.

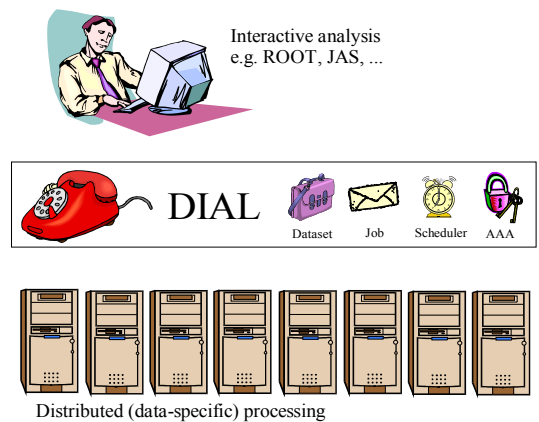

Figure 1. DIAL (center) provides the connection between a user in a interactive analysis environment (top) and distributed processing (bottom).

### 4.1. Front and back ends

There are a number of interactive analysis environments that have been developed inside and outside the context of HEP. Examples within HEP include PAW [6], ROOT [7] and JAS [8]. These systems often define a native data format (such as the PAW ntuple or ROOT tree) and experiments wishing to use the environment traditionally store summary data in that format to provide easy access to the data from the analysis framework.

However, the bulk of an experiment's data is often in one or more other formats that are more natural to the applications used to simulate or reconstruct the data. The design of DIAL explicitly recognizes the existence of these different formats and allows the back-end application used to process the data to differ from the front-end analysis framework. Of course a user has the option of using the same application for both front and back ends.

When the processing is distributed, it is the back-end application that is replicated in multiple jobs. The user interacts with a single instance of the front-end framework and the parallel nature of the back-end processing is hidden.

### 4.2. Major components

Figure 2 shows the interactions between the major components of DIAL. A user sitting inside an analysis framework identifies a scheduler that will be used to control the processing. The user then identifies the back-end application, the processing task to be carried out on each event, and the dataset to which it is to be applied. The user submits the application, task and dataset to the scheduler which splits the dataset along event boundaries and creates and runs a job for each sub-dataset. Each job produces a result and the scheduler concatenates these results and returns the overall result to the user.

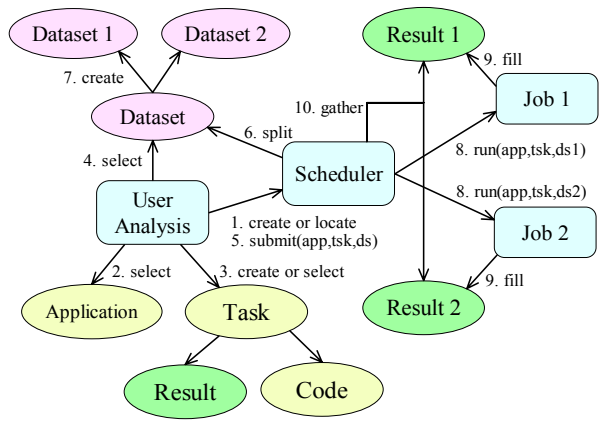

Figure 2. DIAL components and their interactions.

More detail on each of the components is provided in the following sections.

### 4.3. Application

The application specifies the back-end executable used to process the data. This specification includes the name, version and possibly a list of shared libraries. It is the responsibility of the scheduler to locate the appropriate executable and libraries on the machine(s) where the jobs are run. In the present implementation, ChildScheduler, the data required to make this mapping is found in files on the target node. The files are found in locations specified by a combination of environment variables and application name and version.

The application executable provides a loop over events and applies a user-supplied task to each event. The user typically specifies this task as a snippet of programming code, for example in C++. The scheduler is responsible for compiling this code and linking it into the executable. In the same present implementation, the instructions for compiling and linking are again found on the target node indexed by application. The code and its library are also stored there with a similar indexing convention.

### 4.4. Result

The data produced during processing that is returned to the user is called the result. The user specification for a job includes an empty result which is a collection of named empty products. These products are analysis objects which are filled each time an event is processed. An event selection (list of selected events), a histogram and an ntuple are examples of products.

Typically, each sub-job is started with an empty result which is filled when the job is run and the results from the sub-jobs are concatenated to create to overall result. This is done by independently concatenating each of the products in the result. It is conceivable that a sub-job to could be started with a result that has already been filled by one or more other sub-jobs or that a result that could be shared by more than one job. The management of jobs and results including concatenation to form the overall result is the responsibility of the scheduler.

### 4.5. Task

Before a job is submitted, the empty result and the code required to fill the result are bundled together to form a task. The code must be written in a language that is suitable for the back-end application.

### 4.6. Dataset

The dataset specifies the input data for processing, i.e. which events and which data for each event. This very important component is discussed in detail in a later section.

### 4.7. Job

A job is specified by a task, an application to run the task and an input dataset. A job has a status indicating whether and when it was started or stopped, how many events have been processed and its result. It may be possible to obtain a partial result while the job is being processed.

A job is submitted to a scheduler and it is that scheduler that is queried to obtain the status of or result from a job.

### 4.8. Scheduler

The scheduler is the heart of DIAL. Users submit jobs to a scheduler which may run the job directly or pass the request on to another scheduler. The scheduler may divide the dataset into sub-datasets, create a new job for each of these and then run or submit each new job. The scheduler provides means to submit a job, kill a running job and query the status and result for a running or completed job.

Although the user typically interacts with a single scheduler, we envision a hierarchy of schedulers corresponding to the hierarchy of computing resources over which the job may be distributed. One possible hierarchy is illustrated in figure 3 which show a grid scheduler using site schedulers which use farm schedulers which use node schedulers. The latter have the responsibility of running the job with the application executable. A user can enter this hierarchy at any level: submitting a job to a single node, to the grid or to any level in between.

This hierarchy of schedulers makes it possible to distribute jobs over compute nodes that may not be visible to the scheduler receiving the original submission. Client and server schedulers are envisioned to provide the means to pass requests over a network.

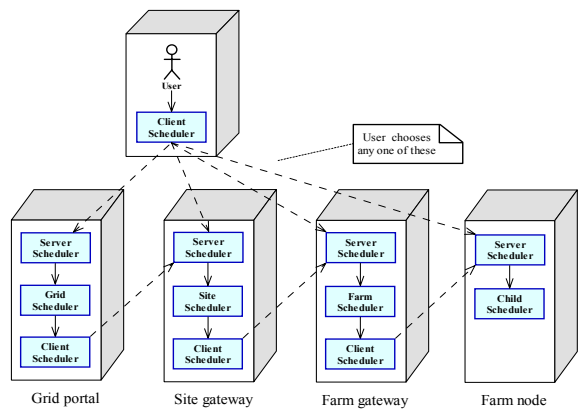

Figure 3. Scheduler hierarchy.

### 4.9. Exchange format

Schedulers communicate by exchanging DIAL objects including jobs, results, applications, and datasets. It is natural to introduce an exchange format to facilitate communication over a network or between different processes on a compute node. We have adopted XML for this purpose and the XML definitions for our components specify the data associated with each. This makes it possible for the analysis framework, each of the schedulers and the back-end application to be implemented in different programming languages.

## 5. OTHER DESIGN CONSIDERATIONS

There are a number of concepts that are missing from the above discussion and our current design and implementation. We list a few of these for completeness.

### 5.1. Authentication and authorization

Authentication and authorization must be taken into account especially where the processing is to be done at one or more remote sites. Certificates likely are the way to handle this on the grid.

### 5.2. Resource allocation

Schedulers need policy and enforcement for allocating the resources to which they have access. This includes balancing multiple requests from multiple users and negotiating with other systems (e.g. batch queues).

### 5.3. Resource estimation

Allocation requires an estimate of the required resources including CPU cycles, disk space and network bandwidth. These estimates might be constructed by comparing information provided by the application and task with the capabilities of the sites, farms or nodes under consideration.

The information can be obtained or at least refined by processing a subset of the requested events. The requirement of interactive response will probably preclude doing this for every job and appropriate predictive data should be stored and associated with the task and application.

### 5.4. Response time

Interactive will have different meanings for different users and so a mechanism should be provided to enable users to specify the desired response time. This might be a characteristic of the scheduler or, more likely, part of the job submission.

### 5.5. Matchmaking

Implied by all the above is a means to do matchmaking, i.e. decide on the appropriate division of the input dataset and then assign the corresponding sub-jobs to other schedulers or real processes.

### 5.6. Adaptive scheduling

A strategy of simply distributing a dataset over multiple jobs and then waiting for all to complete has the serious drawback that the completion time can be no shorter than the time required to process the slowest job. It must be possible to identify stopped or slow jobs and resubmit them or create an equivalent collection of jobs with different granularity.

At some levels it may be appropriate to start processing with a fraction of the data and then submit new jobs where the old jobs complete. This allows an automatic balancing of resources. The ROOT parallelism system PROOF takes this approach.

In some circumstances, it might be appropriate to use aggressive scheduling where the same data is processed on multiple nodes and results from the slower system are discarded.

## 6. DATASETS

The success of an interactive analysis system depends critically on the organization of the input data. It is often the case that the response time is determined more by the time required to access the data rather than the processing time. In this case, performance can be enhanced by only accessing the data of interest and bringing the application to the data rather than moving the data. In a distributed processing system, this implies that the data itself is distributed.

In the following paragraphs, we identify some of the features required of datasets used as the input for a distributed interactive analysis system like DIAL. Many of the comments are relevant to other types of analysis and to batch-oriented production and selection.

### 6.1. Events

In order to use event parallelism, the dataset must be organized into events or their equivalent. Each event should have a unique identifier (unique at least within the context of the dataset) so that the processing can be tracked. Events that have been processed or selected can be specified by a list or range of event identifiers. There must means to obtain the list of identifiers for the events in a dataset.

## 6.2. Content

An event in a dataset is characterized by its content, i.e. the kind of data that the event holds. In the HEP world, examples include raw data, jets, tracks, electrons, etc. It is often useful to specify the content by more that just type: an event may hold electromagnetic and hadronic jets or a collection of jets with cone size 0.5 and another with the value 0.7.

It is sensible to require that a dataset be consistent, that is that each of its events has the same content. In this case we can speak of the content of the dataset. It may be desirable to relax consistency slightly and allow some content to be missing from some of the events.

Specification of content is important because we should ensure that a dataset has the required content before processing. Also if we recognize that a dataset has content that is not needed, we may be able to speed processing by not replicating that part of the data, not moving it across wide-area or local networks, or not reading it into memory.

## 6.3. Data object location

The preceding sections suggest that a dataset is a collection of events each holding a collection data objects indexed by content. Although this is a sensible logical view, it should not be taken too literally. The natural means to access the data depends on the format in which the data is stored and is not part of the dataset specification. What is required is that there are means to locate the data associated with any included event and content. In particular, it should be possible to efficiently iterate over all the events in the dataset.

## 6.4. Files

For simplicity and because of its relevance to grid processing, we assume the data associated with a dataset resides in a unique collection of files. These may be logical files, i.e. identifiers that can be used with a replica catalog to locate one or more physical instantiations of each file. Each of these files holds some of the data for dataset and all of the files are required to access all the data in the dataset. The uniqueness of this file collection may be lost if object replication is allowed. This is discussed later.

Figure 4 shows an example of a dataset and its mapping to files. Events are along the horizontal axis and content along the vertical.

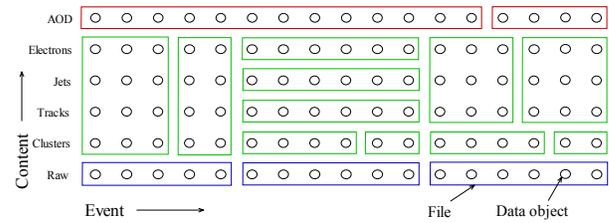

Figure 4. Dataset with mapping to files.

## 6.5. Content selection

We may perform content selection on a dataset, i.e. choose a subset of the content and ask for only the data associated with the restricted content. Clearly the resulting collection of data is also a dataset and the set of files associated with the new dataset is a subset of those associated with the original.

Figure 5 shows the effect of one content selection on the dataset in figure 4. The selected data objects and files are highlighted.

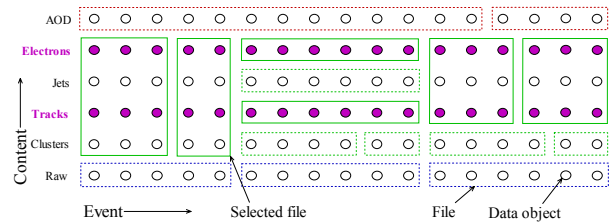

Figure 5. Dataset with content selection.

This example illustrates that content selection can reduce the number of files. It also shows that the extent to which unwanted data is filtered out depends on the original placement of the data. Distributing content over many files can enhance the fraction of data that is useful at the price of increasing the complexity of locating and delivering data when more content is required. It also increases the number of files required to view all the data for a single event.

## 6.6. Event selection

Distributed analysis requires means to divide a dataset along event boundaries, i.e. to do event selection. It is clear again that the data obtained by selecting a subset of the events from dataset is another dataset. Again the files in the new dataset are a subset of those in the original.

Figure 6 shows the effect of an event selection on the content-selected dataset in figure 5. The selected sub-dataset holds two files and only the data of interest.

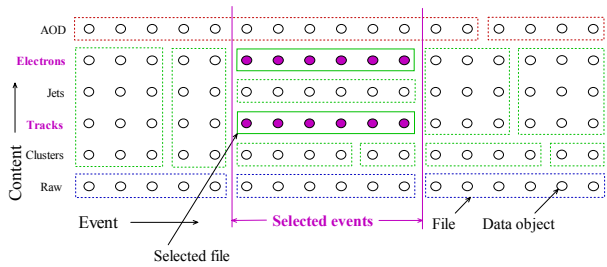

Figure 6. Dataset with content and event selections.

Note that the dataset in this example is split along file boundaries. Clearly this is a desirable feature especially at higher scheduler levels where effort is likely to be expended to move files or locate sites, farms or nodes with the files of interest. At low levels, it may be advantageous to further divide the dataset if file replicas are easily accessed from more than one CPU.

### 6.7. Completeness

It is often the case that the some of the data objects for an event are derived from others in the same event. E.g. tracks are constructed from clusters which are constructed from raw data. We say that a dataset is complete if it includes all the data from which any data object in the event is derived. In our example the content selections raw, raw+cluster and raw+cluster+tracks all correspond to complete datasets. Datasets with content tracks alone, clusters alone or cluster+track are all incomplete.

Typically a user doing analysis is thinking in terms of a complete dataset (the data of interest and the data from which it is derived) but the analysis program only needs access to selected content which is expressed as an incomplete dataset. Results generated by the analysis on the incomplete dataset apply equally well to the complete version. For example, a user might use an incomplete dataset (tracks alone) to select events with one or more tracks meeting some kinematic criteria and then apply this event selection to the corresponding complete dataset (raw+cluster+tracks) to define the input for refitting these tracks.

### 7. DATA PLACEMENT

The dataset view described in the preceding section enables the DIAL scheduler (or the analogous component in any distributed analysis system) to take advantage of the placement of the data. In this section we describe a few data placement strategies that can be employed to improve access time and thus improve interactive response.

### 7.1. Content distribution

Almost any analysis task will only make use of a small fraction of the content in each event. If the content of interest is contained in files distinct from those containing the remainder, then access time (and thus analysis response) time may be greatly improved by ignoring the files with unwanted data.

In its simplest incarnation, this is a standard feature of existing HEP experiments. Data read from the detector (raw data) is processed to create reconstructed data and then summary data is extracted from that. The raw, reconstructed and summary data are stored in different collections of files and often in different formats. The summary data is often in a format natural to the anticipated analysis framework, e.g. PAW or ROOT ntuples. The computing models for the LHC experiments are similar with AOD (Analysis Oriented Data) playing the role of the summary data.

The early plans for most experiments call for keeping a very small amount of summary data for each event so that it can be processed quickly. However there are inevitably users who find the data they need is missing and there is pressure to include more and more information because the complete reconstructed data is difficult to access.

Subdividing the content of both reconstructed and summary data can allow users to more precisely specify the content required for a particular analysis. Distributing the placement of this content over distinct file sets can then allow fast access for users requiring minimal content while still providing easy (but slower) access to users requiring more content.

### 7.2. Exclusive event streaming

Analysis users are only interested in data for a small fraction of the events in the data store. Typically the data of interest are chosen by apply a series of event selection algorithms on a global dataset to obtain increasingly smaller datasets. The selection criteria depend on the physics channel of interest. If the final dataset is small, then the data can be replicated. However, there may be many large datasets for which this is not feasible.

It is desirable to place the data so one can access the data for these large datasets without accessing all the files in the global dataset. This can be accomplished without replication by exclusive event streaming, i.e. identifying selection criteria, assigning exclusive bins in these criteria and then assigning a stream for each bin. Each stream corresponds to a dataset with a distinct set of files. All events meeting the bin criteria are stored in the corresponding dataset.

If the bin criteria are chosen appropriately, then typical analysis datasets can be constructed by merging a small subset of these streams. Larger datasets require more streams.

HEP selection criteria include the acquisition trigger type and reconstructed features such as the missing energy, number of jets and number of leptons. Because of the large number of dimensions, one must be judicious in defining the bins and there may be a "none of the above" category for leftover events.

### 7.3. Object replication

With or without with event streaming, after a few selections, datasets often become sparse, i.e. only a small fraction of the data in the dataset files is included in the dataset. If the dataset is not too large, then it becomes affordable to replicate the included data objects, i.e. copy them to a new set of files.

The new dataset is equivalent to the original in that it has the same events, the same content and the same data for each event and content. It points to a different collection of (logical) files but any analysis performed on the new dataset is expected to give the same result as one carried out on the original.

Note that object replication is distinct from file replication where an entire physical file is replicated but retains its original logical identity. File replication has already been described by introducing logical files and their replica catalog.

One important consideration for object replication is the management of references to data objects. If the reference is expressed as a logical file plus a position in that file, can the replicated object be recognized as satisfying this reference? This can be handled with a global object table but there are serious scaling problems with large numbers of objects as in the LHC event stores.

A more scalable solution is to include a mapping in each file holding replicated objects and consolidate this information in each of the datasets which include this file.

## 7.4. Placement within a file

In addition to limiting the number of files and their sparseness, it is also important that the data within a file be organized in a manner that allows rapid access. Data that are likely to be used together or in succession should be grouped together. Whether to group first by content and then by event or vice versa depends on expected usage patterns. Similarly event iteration should be in a natural access order, e.g. the order in which the events were written, rather than ordered by event identifier.

It is for all these reasons that the dataset does not specify how to access its data, only that there are means to gain access.

## 8. DIAL STATUS

DIAL makes use of another system simply known as dataset [9]. Both have been developed at BNL (Brookhaven National Laboratory). Here we give a brief report on their status.

## 8.1. Existing code

The most recent releases of both DIAL and dataset are numbered 0.20. All the components of DIAL have been implemented but the only scheduler is a simple node scheduler called ChildScheduler that carries out processing as a single job.

The dataset components are also in place but the only concrete implementation of a dataset is the ATLAS AthenaRoot file. These files hold Monte Carlo generator information. Most of ATLAS simulated and reconstructed data can only be stored in zebra files and ATLAS will soon abandon this format.

Both DIAL and dataset are implemented in C++ with methods to read and write XML descriptions of all classes describing object that can be exchanged between DIAL schedulers.

## 8.2. Front end

ROOT may be used as the front-end analysis framework for DIAL. All DIAL and dataset classes have been imported into ROOT using the ROOT ACLiC facility. This means that all DIAL and dataset classes and functions can be accessed from the ROOT command prompt. A script in the package dial_root can be used to fill the ROOT dictionary and load the DIAL and dataset libraries. Only preliminary testing has been done.

In order for ROOT histograms to be used as DIAL analysis objects (i.e. to be included as products in results), it is necessary to add the corresponding DIAL adapters. This has not been done yet but would require only a small amount of effort.

## 8.3. Back end

The only back-end application that has been integrated into DIAL is the test program dialproc which is included as a part of the DIAL release. This is the major missing piece that must be provided to make DIAL a useful tool for local processing. ATLAS candidates for back-end applications are PAW and ROOT to view summary data and athena [10] for reconstructed data.

## 9. DEVELOPMENT PLANS

Here we list some future activities that will enable DIAL to become a useful tool both for studying distributed analysis and for distributed analysis of ATLAS data.

## 9.1. Client-server schedulers

The present expectation in DIAL is that schedulers will carry out most of the network using a client-server mechanism. A server scheduler manages and provides access to one or more destination schedulers on its compute node. This server might run as a daemon or web service. A client scheduler embedded in a source application (analysis framework or scheduler) on another machine would provide the means to communicate with the server over the network. This is illustrated in figure 3.

Implementing this functionality will enable network communication and thus be an important step in building a distributed system.

## 9.2. Farm scheduler

The farm scheduler will make use of the client-server mechanism to communicate with node schedulers on a collection of farm nodes. It will split an input dataset and distribute processing over multiple nodes.

### 9.3. Site and grid schedulers

Further in the future is the development of site and grid schedulers to complete the hierarchy shown in figure 3. These will bring up many new issues including authentication, authorization, dataset and file catalogs, and other grid integration issues.

### 9.4. Concrete datasets

Useful analysis will require that DIAL be able to access interesting data. Starting this summer, ATLAS event data will be stored in LCG POOL [11] event collections. Expressing these collections in terms of datasets will both provide access to ATLAS data and provide important feedback about the evolving POOL collection model.

ATLAS summary data is presently stored in PAW ntuples. We may use these to construct datasets or convert them to ROOT ntuples and then provide a dataset interface for the latter.

### 9.5. Back-end

The natural way to access ATLAS reconstructed data is with athena, the event-loop framework used to create the data. Athena will be made available as a back-end application for DIAL with a special algorithm to call the DIAL code used to fill an analysis result. Event data in athena is normally found in a transient data store called StoreGate. Summary data might be accessed by finding a mechanism to put it in StoreGate or by reading it directly from the PAW or ROOT ntuples.

Either PAW or ROOT may also be implemented as a back-end application for summary data. This would require embedding an event loop in the application.

### 9.6. Front end

As indicated earlier, ROOT has already been included as a front-end analysis framework for DIAL. The LCG SEAL project [12] is developing a Python-based interactive framework that might eventually include analysis tools and thus be a sensible candidate for use with DIAL. We will wait and see how this project evolves.

JAS is a java-based analysis framework that is another possible candidate. Extra effort would be required to provide a java implementation or binding to the existing DIAL code written in C++.

### 9.7. ATLAS

The above plans cover different realms and time scales. The items required to deliver a useful tool for ATLAS are the client-server schedulers, the farm scheduler, a dataset interface to ATLAS POOL collections and integration of an athena back end.

## 10. GRID COMPONENTS AND SERVICES

One of the important goals of DIAL is to provide a tool that can be used to evaluate the requirements that interactive analysis places on grid computing. This includes identification of components and services that can be shared with other distributed analysis systems (such as PROOF and JAS), the distributed batch systems being developed by HEP experiments (such as ATLAS) and non-HEP event-oriented processing systems.

Some possible candidates for shared components and services follow. DIAL is developing the minimum needed to deliver an end-to-end system but would prefer to incorporate shared solutions where available.

### 10.1. Dataset

The desires to use event parallelism and optimize data access lead to the dataset model described earlier. The essential features include organization into event and content with a mapping to logical files. Means are provided to select both content and events.

### 10.2. Job

An analysis user expresses a query or request for data in the form of a job. Parallelism is obtained by splitting the job along event boundaries into sub-jobs until the desired response time can be obtained. The specification of the job includes the input dataset, the application, and the task to run with the application.

### 10.3. Application

An application specifies the executable used to process a job and the required runtime environment including shared libraries. There must be a means to locate the executable and libraries on whatever target node is selected to process the job. If the executable and environment are not available, there might be an automatic system to install them.

### 10.4. Result

The output of a job is a result, i.e. the result carries the response to the user query. It must be possible to merge the results from sub-jobs to obtain the overall result. The result is a collection of analysis data objects. The AIDA (Abstract Interfaces for Data Analysis) [13] project has proposed a standard interface for analysis classes. These might be relevant.

### 10.5. Scheduler

A scheduler is a service that accepts a job submission and makes use of other services to locate computing resources and input data and then do matchmaking between these. It divides the job into sub-jobs and then runs these jobs or submits them to another scheduler. It gathers and concatenates results when these jobs complete.

The scheduler must also monitor the progress of jobs and take corrective action if jobs fail or progress slowly. And it provides means for users to query the status of existing jobs including concatenating and returning partial results.

## 10.6. Authentication and authorization

There must be means for users to identify themselves so that their requests can be authorized. This authentication must be distributed along with the job processing so that authorization and allocation can be performed at each level.

## 10.7. Resource location and allocation

The scheduler will make use resource location services to find computing resources and then negotiate with allocation services to gain access to the required compute resources. This will be done in competition with other DIAL schedulers and with schedulers or their equivalent for other systems, both interactive and batch.

## 11. INTERACTIVE GRID

The above components and services are not unique to interactive analysis but are also needed for data production and batch-oriented analysis. Here we discuss some requirements that are most relevantt to an interactive system.

### 11.1. Latency

The fundamental difference between batch and interactive systems is the low latency required by the latter. An interactive system should provide the user with a means to specify the maximum acceptable response time and then distribute the job in a manner that satisfies this constraint. This time includes the time to locate data and resources, do matchmaking, split datasets and jobs, submit jobs, execute these jobs, and gather and concatenate results.

### 11.2. Job caching

An analysis user often processes the same dataset many times in succession. A user may be willing to tolerate a slow response to the first request to process a dataset because of the time required to locate data and resources, determine how to distribute the request, and move data that is not already in place. In subsequent passes over the same dataset, the user reasonably expects the scheduler to remember and make use of the configuration from earlier passes.

### 11.3. Dynamic job scheduling

An interactive scheduler must closely monitor the progress of its jobs. Like a batch system, it must recognize and resubmit when a job fails or hangs. An interactive scheduler should closely track the progress of its jobs and verify that they will complete within the specified response time. If not, the scheduler might request more resources for those jobs, move them to different location, or subdivide and resubmit.

### 11.4. Resource allocation

Interactive response imposes requirements on resource allocation. In addition to requesting a number of CPU cycles, one needs to request that these be delivered in a specified clock time. This may require restricting the search to fast processors or those that are lightly loaded. It may require guaranteeing that additional loads are not imposed on the processor while the interactive request is active. The interactive job may have to be pushed to the front of a queue ore even preempt running jobs. A reservation system might be imposed to guarantee the desired response.

### 11.5. Progress and partial results

Inevitably the analysis user will make requests which cannot be fully processed in a time deemed to be interactive. In this case the user will want to monitor the progress of the job and know the fraction of events processed, the estimated time to complete and to obtain a partial result, i.e. some or all of the result for those events that have been processed thus far.

## 12. SHARING RESOURCES

Of course many of the requirements imposed by an interactive system could be met by providing interactive users (or, even better, each interactive user) with dedicated resources. However there is considerable fluctuation in the demand for interactive analysis depending on the time of day, time to the next conference and the discovery of an interesting data feature. Thus, at least for analysis of large datasets, it is likely desirable to share resources with activities such as data reconstruction and Monte Carlo production that require many CPU cycles but do not impose strict constraints on latency.

## 13. CONCLUSIONS

The DIAL project has been established to demonstrate the feasibility of interactively analyzing large datasets, to set requirements for grid components and services, and to provide ATLAS with a useful distributed analysis system.

The design allows a user to independently select an analysis framework and a data-processing application and then carry out interactive analysis by submitting a series of jobs to a DIAL scheduler. Each job produces a result which is a collection of analysis objects that can be manipulated (displayed, fitted, etc.) in the analysis framework. DIAL will provide a hierarchy of schedulers to distribute the processing in a transparent manner. An important component of this design is the data view provided by datasets.

We expect the grid to deliver the compute resources required to access and process very large datasets. Some of the relevant grid components and services have been

*CHEP03, UCSD, March 24-28 2003* 10CHEP03, UCSD, March 24-28 2003 10

described with emphasis on the requirements imposed by an interactive analysis system. DIAL will continue to identify these components and services and will provide schedulers for farm, site and grid operation.

## Acknowledgments

The author wishes to thank members of the ATLAS database group, the PPDG interactive analysis group, the LGG POOL group and the PAS (Physics Application Software) group at BNL for many useful discussions that have led to the formulation of the design described here.

This research is supported by the LDRD (Laboratory Directed Research and Development) program at BNL and by the ATLAS and PPDG projects at BNL through the Division of High Energy Physics under DOE Contract No. DE-AC02-98CH10886.